\documentclass[aps,preprint,showpacs]{revtex4}

\usepackage{amsmath}
\usepackage{amsfonts}
\usepackage{amssymb}
\usepackage{graphicx}

\begin{document}

\title{\textbf{Quantum Resonances and Ratchets in Free-Falling Frames}}
\author{\textbf{Itzhack Dana and Vladislav Roitberg}}
\affiliation{Minerva Center and Department of Physics, Bar-Ilan University,
Ramat-Gan 52900, Israel}

\begin{abstract}
Quantum resonance (QR) is defined in the free-falling frame of the quantum
kicked particle subjected to gravity. The general QR conditions are derived.
They imply the rationality of the gravity parameter $\eta $, the
kicking-period parameter $\tau /(2\pi )$, and the quasimomentum $\beta $.
Exact results are obtained concerning wave-packet evolution for arbitrary
periodic kicking potentials in the case of integer $\tau /(2\pi )$ (the 
main QRs). It is shown that a quantum ratchet generally arises in this 
case for resonant $\beta $. The noninertial nature of the free-falling frame 
affects the ratchet by effectively changing the kicking potential to one 
depending on $(\beta ,\eta )$. For a simple class of initial wave packets, 
it is explicitly shown that the ratchet characteristics are determined to 
a large extent by symmetry properties and by number-theoretical features 
of $\eta $.\newline
\end{abstract}

\pacs{05.45.Mt, 05.45.Ac, 03.65.-w, 05.60.Gg}
\maketitle

The quantum kicked particle in the presence of gravity has attracted much
interest recently \cite{ao,ao1,ao2,fgr,qam,qam1,qam2} following the
experimental discovery of the ``quantum accelerator modes'' (QAMs) of freely
falling atoms periodically kicked by pulses \cite{ao}. The QAMs were
observed in the free-falling frame in a strong quantum regime and their
explanation was at the focus of all the theoretical studies \cite
{fgr,qam,qam1,qam2} which we briefly summarize here. Using dimensionless
quantities and notation introduced in Ref. \cite{fgr}, the general
Hamiltonian for the system is 
\begin{equation}
\hat{H}=\frac{\hat{p}^{2}}{2}-\frac{\eta }{\tau }\hat{x}+kV(\hat{x}
)\sum_{t}\delta (t^{\prime }-t\tau ),  \label{H}
\end{equation}
where $(\hat{p},\hat{x})$ are momentum and position operators, $\eta $ is
proportional to the gravity force (in the direction of the positive $x$
axis), $\tau $ is the kicking period, $k$ is a nonintegrability parameter, $
V(x)$ is a periodic potential, and $t^{\prime }$ and $t$ are the continuous
and ``integer'' times. The units are chosen so that the particle mass is $1$, 
$\hbar =1$, and the period of $V(x)$ is $2\pi $; the standard potential $
V(x)=\cos (x)$ was used in all works. The transformation to the free-falling
frame is accomplished by applying the gauge transformation $\exp (i\eta \hat{
x}t^{\prime }/\tau )$ to the Schr\"{o}dinger equation for (\ref{H}) \cite
{fgr}. One then finds that the Hamiltonian in this frame is 
\begin{equation}
\hat{H}_{\text{f}}=\frac{\left( \hat{p}+\eta t^{\prime }/\tau \right) ^{2}}{2
}+kV(\hat{x})\sum_{t}\delta (t^{\prime }-t\tau ).  \label{Hf}
\end{equation}
Unlike (\ref{H}), $\hat{H}_{\text{f}}$ in (\ref{Hf}) is translationally
invariant in $\hat{x}$, implying the conservation of a quasimomentum $\beta $
(the ``fractional part'' of $\hat{p}$, $0\leq \beta <1$) in the quantum
evolution under $\hat{H}_{\text{f}}$; at fixed $\beta $, one can consider $x$
as an angle $\theta $ and write $\hat{p}=$ $\hat{N}+\beta $, where $\hat{N}
=-id/d\theta $ is an angular-momentum operator with integer eigenvalues $n$
(see more details in note \cite{note}). Then, (\ref{Hf}) becomes the
Hamiltonian of a kicked-rotor system in the free-falling frame. The
one-period evolution operator, from\ $t^{\prime }=t\tau +0$ to $t^{\prime
}=(t+1)\tau +0$, is given by 
\begin{equation}
\hat{U}_{\beta }(t)=\exp \left[ -ikV(\hat{\theta})\right] \exp \left[
-i(\tau /2)\left( \hat{N}+\beta +\eta t+\eta /2\right) ^{2}\right] ,
\label{Ubt}
\end{equation}
up to an irrelevant constant phase factor. Now, the value of $\tau =2\pi
l_{0}$, where $l_{0}$ is a positive integer, corresponds to the main quantum
resonances (QRs) of (\ref{Ubt}) in the absence of gravity ($\eta =0$) \cite
{kp,dd,dd1}. It was shown in Ref. \cite{fgr} that $\tau =2\pi l_{0}+\epsilon 
$\ defines, for sufficiently small $\epsilon \neq 0$ and for any $\eta $, a
``quasiclassical'' regime in which $\epsilon $ plays the role of a
fictitious Planck's constant. In this regime, the quantum evolution under
(\ref{Ubt}) can be approximately described by a classical map. Then, a wave
packet initially trapped in an accelerator-mode island of this map
``accelerates'', i.e., the expectation value $\left\langle \hat{N}
\right\rangle $ of $\hat{N}$ in the wave packet grows linearly in time; this
is a QAM. The experimentally observed robustness of QAMs under variations of 
$\tau $ near $\tau =2\pi l_{0}$ was explained, in the framework of the
quasiclassical approximation, as a ``mode-locking'' phenomenon \cite
{qam,qam1,qam2} (see also conclusion). Theoretical predictions were verified
by several experiments \cite{ao2,qam}.\newline

In this paper, the system (\ref{Ubt}) with arbitrary periodic potential $
V(\theta )$ is systematically approached in a different way. The concept
of QR is introduced for this system ($\eta \neq 0$) and exact results are
derived concerning its quantum-resonant dynamics. A consistent definition
of QR for $\eta \neq 0$ requires the time-dependent operator (\ref{Ubt})
to be essentially periodic in $t$ with some finite period $T$; QR can then
be defined on the basis of the evolution operator in $T$ kicks. The
general conditions for QR, given by Eqs. (\ref{Om})-(\ref{b}) below, imply
the rationality of $\eta $, $\tau /(2\pi )$, and $\beta $. Exact results
for wave-packet evolution under (\ref{Ubt}) are obtained in the case of
integer $\tau /(2\pi )$ (main QRs). We find that in this case the
noninertial nature of the free-falling frame effectively changes $V(\theta 
)$ to a potential $V_{\beta ,\eta }(\theta )$. We then show that a linear 
growth of $\left\langle \hat{N}\right\rangle$ in time generally occurs for 
resonant quasimomentum (\ref{b}). This is a purely quantum ``ratchet'' effect 
\cite{qra,qrr}, a directed current without a biased force, caused, e.g., by 
some asymmetry in the system. QR ratchets have been investigated recently 
\cite{qrr} for the usual kicked rotor ($\beta =\eta =0$). We emphasize that 
for $\eta \neq 0$ there is {\em no} biased force in the system (\ref{Hf}): 
gravity is classically not felt in the free-falling frame. In fact, (\ref{Hf}) 
satisfies the conditions for a ratchet Hamiltonian \cite{qra} but the kinetic 
energy is time dependent, reflecting the noninertial nature of the free-falling
frame. It is this time dependence that affects the QR ratchet through the 
effective potential $V_{\beta ,\eta }(\theta )$. In particular, the ratchet
current vanishes if $V_{\beta ,\eta }(\theta )$ and the initial wave packet 
have a common point symmetry. For a simple class of initial wave packets, we 
derive closed explicit results for the linear-growth coefficient and we show 
that the ratchet characteristics are determined to a large extent by symmetry 
properties and by number-theoretical features of $\eta $.\newline

{\em QR in free-falling frames.} QR is the quadratic growth of the
kinetic-energy expectation value in time, due to the translational
invariance of some basic evolution operator $\hat{U}$ for the system in
phase space; this invariance leads to a band quasienergy spectrum of $\hat{U}
$ and thus to QR. A basic operator $\hat{U}$ for (\ref{Ubt}) can be
consistently defined only if $\hat{U}_{\beta }(t+T)=\hat{U}_{\beta }(t)$ for
some period $T$. Then $\hat{U}=\hat{U}_{\beta ,T}(t)$, where 
\begin{equation}
\hat{U}_{\beta ,T}(t)=\hat{U}_{\beta }(t+T-1)\cdots \hat{U}_{\beta }(t+1)
\hat{U}_{\beta }(t)  \label{UbT}
\end{equation}
is the evolution operator in $T$ kicks, and the operators $\hat{U}_{\beta
,T}(t)$ for all $t$ are equivalent (similar), due to $\hat{U}_{\beta
,T}(t+1)=\hat{U}_{\beta }(t)\hat{U}_{\beta ,T}(t)\hat{U}_{\beta }^{-1}(t)$.
This allows one to associate with $\hat{U}_{\beta ,T}(t)$ a meaningful
(essentially $t$-independent) quasienergy problem. To derive explicit
conditions for $\hat{U}_{\beta }(t+T)=\hat{U}_{\beta }(t)$, one must exploit
the fact that $\hat{U}_{\beta }(t)$ is defined up to an arbitrary,
physically irrelevant phase factor which may depend only on $t$. We thus
replace $\hat{U}_{\beta }(t)$ by $\hat{U}_{\beta }^{\prime }(t)=\exp
(ia_{1}t+ia_{2}t^{2})\hat{U}_{\beta }(t)$, where $a_{1}$ and $a_{2}$ are
constants to be determined. Using (\ref{Ubt}) and the fact that $\hat{N}$
has integer eigenvalues, we easily get from $\hat{U}_{\beta }^{\prime }(t+T)=
\hat{U}_{\beta }^{\prime }(t)$ that $a_{1}=\tau \eta (\beta +\eta /2)$, $
a_{2}=\tau \eta ^{2}/2$, and 
\begin{equation}
\Omega \equiv \frac{\tau \eta }{2\pi }=\frac{w}{T},  \label{Om}
\end{equation}
where $w$ is some integer. Eq. (\ref{Om}), i.e., the rationality of $\Omega$,
is the only physically relevant condition for the existence of a basic
operator $\hat{U}_{\beta ,T}(t)$. We shall assume that $(w,T)$ are coprime,
so that $T$ is the smallest period for given rational value of $\Omega $. 
\newline

We now require $\hat{U}_{\beta ,T}(t)$ to satisfy the basic QR condition for
kicked-rotor systems \cite{dd,qr,cs}, i.e., to be invariant under
translations $\hat{T}_{q}=\exp (-iq\hat{\theta})$ by $q$\ (an integer) in
the angular momentum $\hat{N}$: $[\hat{U}_{\beta ,T}(t),\hat{T}_{q}]=0$. In
the last relation, we can neglect, of course, any $t$-dependent phase factor
attached to $\hat{U}_{\beta }(t)$ (see above) and just use the definition 
(\ref{UbT}) of $\hat{U}_{\beta ,T}(t)$ with $\hat{U}_{\beta }(t)$ given by
(\ref{Ubt}). Using also (\ref{Om}) and, again, the fact that $\hat{N}$ has
integer eigenvalues, we find after a straightforward calculation that $[\hat{
U}_{\beta ,T}(t),\hat{T}_{q}]=0$ implies that 
\begin{equation}
\frac{\tau }{2\pi }=\frac{l}{q},  \label{QR}
\end{equation}
\begin{equation}
\beta =\frac{r}{lT}-\frac{q}{2}-\frac{qw}{2l}\ \ {\rm mod}(1),  \label{b}
\end{equation}
where $l$ and $r$ are integers. The QR conditions (\ref{QR}) and (\ref{b})
can be analyzed as in the $\eta =0$ case \cite{dd}. Assuming, for
definiteness and without loss of generality, that $l$ and $q$ are positive,
we write $l=gl_{0}$ and $q=gq_{0}$, where $l_{0}$ and $q_{0}$ are coprime
positive integers and $g$ is the greatest common factor of $(l,q)$. It is
then clear that already at fixed $\tau /(2\pi )=l_{0}/q_{0}$ a resonant
quasimomentum (\ref{b}) can take {\em any} rational value $\beta _{{\rm r}}$
in $[0,1)$; this is because $g$ can be always chosen so that $r=[\beta _{
{\rm r}}+gq_{0}/2+wq_{0}/(2l_{0})]gl_{0}T$ is integer. For given $\beta
=\beta _{{\rm r}}$, we shall choose $g$ as the smallest positive integer
satisfying the latter requirement, so as to yield the minimal values of $
l=gl_{0}$ and $q=gq_{0}$. We denote $\beta _{{\rm r}}$ by $\beta _{r,g}$,
where the integer $r$ above labels all the different values of $\beta _{{\rm
r}}$ for given minimal $g$.\newline

The quasienergy states $\phi (\theta )$ for $\beta =\beta _{r,g}$ are the
simultaneous eigenstates of $\hat{U}_{\beta ,T}(t)$ and $\hat{T}_{q}$: $
\hat{U}_{\beta ,T}(t)\phi (\theta )=\exp (-i\omega )\phi (\theta )$, $\hat{T}
_{q}\phi (\theta )=\exp (-iq\alpha )\phi (\theta )$, where $\omega $ is the
quasienergy and $\alpha $ is a ``quasiangle'', $0\leq \alpha <2\pi /q$.
Using standard methods \cite{qr,cs}, it is easy to show that at fixed $
\alpha $ one generally has $q$ quasienergy levels $\omega _{b}(\alpha ,\beta
)$, $b=0,\dots ,\ q-1$; as $\alpha $ is varied continuously, these $q$
levels typically ``broaden'' into $q$ distinct bands (having nonzero width).
This leads to QR, i.e., the asymptotic behavior $\left\langle \psi _{vT}|
\hat{N}^{2}|\psi _{vT}\right\rangle \sim 2D(vT)^{2}$; here $v$ is a large
integer, $\psi _{vT}(\theta )=\hat{U}_{\beta ,T}^{v}(0)\psi _{0}(\theta )$
is any evolving wave packet, and $D$ is some coefficient.\newline

{\em Case of main QRs.} From now on, we shall focus on the case of $\tau
=2\pi l_{0}$ ($q_{0}=1$), the main QRs. The quantum evolution of wave
packets under (\ref{Ubt}) can be exactly calculated in this case for
arbitrary values of $\eta $ and $\beta $, i.e., not just for the QR values
determined by Eqs. (\ref{Om}) and (\ref{b}). In fact, since $\hat{N}$ has
integer eigenvalues, the relation $\exp \left( -i\pi l_{0}\hat{N}^{2}\right)
=$ $\exp \left( -i\pi l_{0}\hat{N}\right) $ holds, so that (\ref{Ubt}) can
be expressed for $\tau =2\pi l_{0}$ as follows: 
\begin{equation}
\hat{U}_{\beta }(t)=\exp \left[ -ikV(\hat{\theta})\right] \exp \left[
-i\left( \tau _{\beta }+\pi l_{0}\eta +2\pi l_{0}\eta t\right) \hat{N}\right]
,  \label{Ubt1}
\end{equation}
where $\tau _{\beta }=\pi l_{0}(2\beta +1)$ and an irrelevant phase factor
has been neglected. We note that the second exponential operator in (\ref
{Ubt1}) is just a shift in $\theta $. Thus, the result of successive
applications of (\ref{Ubt1}) on an initial wave packet $\psi _{0}(\theta )$
can be written in a closed form: 
\begin{equation}
\psi _{t}(\theta )=\hat{U}_{\beta }(t-1)\cdots \hat{U}_{\beta }(1)\hat{U}
_{\beta }(0)\psi _{0}(\theta )=\exp \left[ -ik\bar{V}_{\beta ,\eta
,t}(\theta )\right] \psi _{0}(\theta -\tau _{\beta }t-\pi l_{0}\eta t^{2}),
\label{pbt}
\end{equation}
where 
\begin{equation}
\bar{V}_{\beta ,\eta ,t}(\theta )=\sum_{s=0}^{t-1}V\left( \theta -\tau
_{\beta }s-2\pi l_{0}\eta ts+\pi l_{0}\eta s^{2}\right) .  \label{Vbet}
\end{equation}

More explicit expressions for (\ref{pbt}) and (\ref{Vbet}) can be obtained
for $\eta =w/(l_{0}T)$, i.e., the values of $\eta $\ corresponding to the
main QRs ($\tau =2\pi l_{0}$) by Eq. (\ref{Om}). Let us leave $\beta $
arbitrary for the moment and choose the time $t$ in a natural way as a
multiple $v$ of the basic period $T$, $t=vT$. Then, writing $s=s^{\prime
}+v^{\prime }T$, with $s^{\prime }=0,\dots ,T-1$ and $v^{\prime }=0,\dots
,v-1$, the sum in (\ref{Vbet}) can be decomposed into two sums over $
s^{\prime }$ and $v^{\prime }$. Using also the Fourier expansion 
\begin{equation}
V(\theta )=\sum_{m}V_{m}\exp (-im\theta ),  \label{V}
\end{equation}
we find from Eqs. (\ref{pbt}) and (\ref{Vbet}) with $\eta =w/(l_{0}T)$ that 
\begin{equation}
\psi _{vT}(\theta )=\exp \left[ -ik\bar{V}_{\beta ,\eta ,vT}(\theta )\right]
\psi _{0}\left( \theta -\tau _{\beta ,w}vT\right) ,  \label{pbet}
\end{equation}
where $\tau _{\beta ,w}=\tau _{\beta }+\pi w=\pi (2l_{0}\beta +l_{0}+w)$ and 
\begin{equation}
\bar{V}_{\beta ,\eta ,vT}(\theta )=\sum_{m}V_{m}W_{m,\beta ,\eta }\frac{\sin
(m\tau _{\beta ,w}vT/2)}{\sin (m\tau _{\beta ,w}T/2)}e^{im(v-1)\tau _{\beta
,w}T/2}\exp (-im\theta ).  \label{VbeT}
\end{equation}
Here 
\begin{equation}
W_{m,\beta ,\eta }=\sum_{s=0}^{T-1}\exp \left[ im\left( \tau _{\beta }s-\pi
ws^{2}/T\right) \right]  \label{ff}
\end{equation}
is a ``form factor'' reflecting the noninertial nature of the free-falling
frame, i.e., the time dependence of the kinetic energy in (\ref{Hf}), in one
period $T$. This factor, which is a generalized Gauss sum \cite{qam1,gs},
effectively changes $V_{m}$ in (\ref{VbeT}) to $V_{m}W_{m,\beta ,\eta }$,
which may be considered as the harmonics of a potential $V_{\beta ,\eta
}(\theta )=\sum_{m}V_{m}W_{m,\beta ,\eta }\exp (-im\theta )$. For $T=1$,
corresponding to $\eta =w/l_{0}$ (and, of course, also to $\eta =0$), $
W_{m,\beta ,\eta }=1$ and $V_{\beta ,\eta }(\theta )=V(\theta )$. Then, the
only effect of gravity on (\ref{pbet}) is through the quantity $\tau _{\beta
,w}$.\newline

{\em QR ratchets}. The general QR behavior $\left\langle \psi _{vT}|\hat{N}
^{2}|\psi _{vT}\right\rangle \sim 2D(vT)^{2}$ for resonant $\beta =\beta
_{r,g}$ (see above) suggests that a quantum-ratchet effect, i.e., a linear growth
of $\left\langle \hat{N}\right\rangle _{vT}\equiv \left\langle \psi _{vT}|
\hat{N}|\psi _{vT}\right\rangle $ under the evolution (\ref{pbet}), may also
occur for $\beta =\beta _{r,g}$ and sufficiently large $v$: 
\begin{equation}
\left\langle \hat{N}\right\rangle _{vT}\approx \left\langle \hat{N}
\right\rangle _{0}+RvT,  \label{R}
\end{equation}
where $R$ is some nonzero coefficient. We now show that this is indeed the
case for general potentials (\ref{V}) and initial wave packets 
$\psi _{0}(\theta )$. At the same time, a formula for $R$ is derived.\newline

We start from the general expansion 
\begin{equation}
\left| \psi _{0}(\theta )\right| ^{2}=\frac{1}{2\pi }\sum_{m}C(m)\exp
(im\theta ),  \label{pbc}
\end{equation}
where $C(m)=\sum_{n}\widetilde{\psi }_{0}(m+n)\widetilde{\psi }_{0}^{\ast
}(n)$ are correlations of the initial wave packet in its angular-momentum
representation $\widetilde{\psi }_{0}(n)$. Using (\ref{pbet}), (\ref{VbeT}),
and (\ref{pbc}), we get 
\begin{eqnarray}
\left\langle \hat{N}\right\rangle _{vT} &=&-i\int_{0}^{2\pi }d\theta \psi
_{vT}^{\ast }(\theta )\frac{d\psi _{vT}(\theta )}{d\theta }  \nonumber \\
&=&\left\langle \hat{N}\right\rangle _{0}+ik\sum_{m\neq 0}mV_{m}W_{m,\beta
,\eta }C(m)\frac{\sin (m\tau _{\beta ,w}vT/2)}{\sin (m\tau _{\beta ,w}T/2)}
e^{-im(v+1)\tau _{\beta ,w}T/2},  \label{Na}
\end{eqnarray}
where normalization of $\psi _{0}(\theta )$ is assumed, $\int_{0}^{2\pi
}\left| \psi _{0}(\theta )\right| ^{2}d\theta =1$. Now, a linear growth of
(\ref{Na}) in $v$ can arise only if $m\tau _{\beta ,w}T/2=r_{m}\pi $ for some 
$m\neq 0$, where $r_{m}$ is integer; then, the contribution of the last 
three terms in which $T$ appears in (\ref{Na}) is just equal to $v$. Using 
$\tau _{\beta ,w}=\pi (2l_{0}\beta +l_{0}+w)$ in $m\tau _{\beta ,w}T/2=r_{m}
\pi $, we find that $\beta $ must satisfy 
\begin{equation}
\beta =\frac{r_{m}}{ml_{0}T}-\frac{1}{2}-\frac{w}{2l_{0}}\ \ {\rm mod}(1).
\label{bm}
\end{equation}
By comparing (\ref{bm}) with Eq. (\ref{b}), in which $l=gl_{0}$ and $
q=gq_{0}=g$ for some ``minimal'' $g$ (see above), we see that (\ref{bm})
gives just a resonant value $\beta _{r,g}$ of $\beta $: $m$ is some multiple
of $g$ ($m=jg$, $j$ integer) and $r_{m}=j[r+l_{0}Tg(1-g)/2]$ for some
integer $r$. Then, by collecting all the terms with $m=jg$ in (\ref{Na}), we
obtain a formula for the coefficient $R$ in (\ref{R}): 
\begin{equation}
R=-\frac{2kg}{T}\sum_{j>0}j{\rm Im}\left[ V_{jg}W_{jg,\beta ,\eta }C(jg)
\right] .  \label{RC}
\end{equation}
Thus, for given resonant quasimomentum $\beta =\beta _{r,g}$, $R\neq 0$ only
if there exist sufficiently high harmonics $V_{m}$ and correlations $C(m)$,\
with $m=jg$, and the sum of the corresponding terms in (\ref{RC}) is
nonzero. These conditions are satisfied by general $V(\theta )$ and $\psi
_{0}(\theta )$. A very simple case of $R=0$ is when $V_{m}W_{m,\beta ,\eta
}C(m)$ is real for all $m$. This occurs, e.g., when the system is
``symmetric'', i.e., when both the effective potential $V_{\beta ,\eta
}(\theta )=\sum_{m}V_{m}W_{m,\beta ,\eta }\exp (-im\theta )$ and $\psi
_{0}(\theta )$ have a point symmetry around the {\em same} center, say $
\theta =0$: $V_{\beta ,\eta }(-\theta )=V_{\beta ,\eta }(\theta )$ and $\psi
_{0}(-\theta )=\pm \psi _{0}(\theta )$ (inversion) or $\psi _{0}^{\ast
}(-\theta )=\pm \psi _{0}(\theta )$ (inversion with time reversal); this
implies that $V_{m}W_{m,\beta ,\eta }$ and $C(m)$ [see (\ref{pbc})] are both
real. We emphasize that the QR quadratic behavior of $\left\langle \psi
_{vT}|\hat{N}^{2}|\psi _{vT}\right\rangle $ is usually {\em not} affected by
such symmetries (see example below).\newline

As an illustration, we consider the simple class of initial wave packets $
\psi _{0}(\theta )=F\left[ 1+A\exp (-i\theta )\right] $, where $A$ is some
complex constant and $F=\left[ 2\pi \left( 1+|A|^{2}\right) \right] ^{-1/2}$
is a normalization factor. Writing $A=|A|\exp (i\gamma )$, we see that $\psi
_{0}(\theta )$ has a symmetry center at $\theta =\gamma $, i.e., $\psi
_{0}^{\ast }(2\gamma -\theta )=\psi _{0}(\theta )$. The only nonzero
correlations in (\ref{pbc}) are $C(0)=1$, $C(1)=2\pi A^{\ast }F^{2}$, and $
C(-1)=C^{\ast }(1)$. Thus, $R\neq 0$ in (\ref{RC}) only for resonant
quasimomenta $\beta =\beta _{r,g}$ with $g=1$; one has $R=-(2k/T){\rm Im}
[V_{1}W_{1,\beta ,\eta }C(1)]$, so that no essential generality is lost by
choosing $V(\theta )=\cos (\theta )$ from now on. To obtain a more explicit
expression for $R$, one has to evaluate $W_{1,\beta ,\eta }$. Let us assume,
for simplicity, that $w$ is positive and odd and $l_{0}=1$, so that $
l=gl_{0}=1$, $q=gq_{0}=1$ and, from Eq. (\ref{b}), $\beta =\beta _{r,1}=r/T$,
$r=0,\dots ,T-1$. For convenience, the latter set of $\beta $ values will
be arranged in a different order, $\beta =\beta _{r,1}=rw/T\ {\rm mod}(1)$
[recall that $(w,T)$ are coprime]. Then, the form factor (\ref{ff}) for $m=1$
and $\tau _{\beta }=\pi l_{0}(2\beta +1)=2\pi rw/T+\pi $ can be exactly
calculated using known results about Gauss sums \cite{gs1}. We find that $
W_{1,\beta ,\eta }=\sqrt{T}J\exp (i\gamma _{\beta ,\eta })$, where the values 
of $J$ and the phase $\gamma _{\beta ,\eta }$ are listed in Table I for three
different cases of $T$.

Table I. Values of $J$ and $\gamma _{\beta ,\eta }$ for $\beta =rw/T\ {\rm 
mod}(1)$, $r=0,\dots ,T-1$. 
\[
\begin{tabular}{||c||c||c||}
\hline\hline
Case & $J$ & $\gamma _{\beta ,\eta }$ \\ \hline\hline
$T$ even & $\sqrt{2}\cos (\pi w/4)\left( \frac{2T}{w}\right) $ & ${\bf \pi }
\left[ r+wr^{2}/T+w(T-1)/4\right] $ \\ \hline\hline
$T\ {\rm mod}(4)=1$ & $\left( \frac{2w}{T}\right) $ & $\pi \left(
r+wr^{2}/T\right) $ \\ \hline\hline
$T\ {\rm mod}(4)=3$ & $\left( \frac{2w}{T}\right) $ & $\pi \left(
r+wr^{2}/T+1/2\right) $ \\ \hline\hline
\end{tabular}
\]
In Table I, $\left( \frac{a}{b}\right) $ denotes the Jacobi symbol \cite{js},
so that $J=\pm 1$. The effective potential is $V_{\beta ,\eta }(\theta )=
\sqrt{T}J\cos (\theta -\gamma _{\beta ,\eta })$ and has a symmetry center at $
\theta =\gamma _{\beta ,\eta }$, i.e., $V_{\beta ,\eta }(2\gamma _{\beta ,\eta
}-\theta )=V_{\beta ,\eta }(\theta )$. We obtain from all the results above: 
\begin{equation}
R=\frac{k|A|J}{\sqrt{T}(1+|A|^{2})}\sin (\gamma -\gamma _{\beta ,\eta }).
\label{Re}
\end{equation}
We thus see from (\ref{Re}) that the noninertial nature of the free-falling
frame causes a phase shift by $\gamma _{\beta ,\eta }$ and a suppression of $R$
by a factor of $\sqrt{T}$, relative to the case when gravity is absent (with 
$T=J=1$ and $\gamma _{\beta ,\eta }=0$). At fixed $|A|$, $k$, and $T$, $|R|$ is
completely determined by the distance $\Delta \gamma =|\gamma -\gamma
_{\beta ,\eta }|$ between the symmetry centers of $\psi _{0}(\theta )$ and $
V_{\beta ,\eta }(\theta )$. For $\Delta \gamma =0$, these centers coincide
and $R=0$; $|R|$ is largest for $\Delta \gamma =\pi /2$, a value which may be
viewed as corresponding to a ``maximal asymmetry'' situation. The ratchet-current
direction is always given by the sign of $J\sin (\gamma -\gamma _{\beta ,\eta })$,
where $J=\pm 1$ depends entirely on number-theoretical features of $(w,T)$
(see Table I). The symmetry properties do not affect the QR quadratic
behavior $\left\langle \psi _{vT}|\hat{N}^{2}|\psi _{vT}\right\rangle \sim
2D(vT)^{2}$. In fact, using $\left\langle \psi _{vT}|\hat{N}^{2}|\psi
_{vT}\right\rangle =$ $\int_{0}^{2\pi }d\theta \left| d\psi _{vT}(\theta
)/d\theta \right| ^{2}$, (\ref{pbet}), and (\ref{VbeT}), we easily find that 
$D=k^{2}/(4T)$, independent of $(\gamma ,\gamma _{\beta ,\eta })$.\newline

In conclusion, QR can be consistently defined for the system (\ref{Ubt})
provided the rationality condition (\ref{Om}) is satisfied. It should be
noted that $\Omega =\tau \eta /(2\pi )$ in (\ref{Om}) is one of the two
parameters featured by the classical map which approximates (\ref{Ubt}) in
the quasiclassical regime of $\tau =2\pi l_{0}+\epsilon $ \cite
{fgr,qam,qam1,qam2} (see also introduction); the second parameter is a
nonintegrability one, $\tilde{k}=k|\epsilon |$. For sufficiently small $
\tilde{k}$, there exist accelerator-mode islands whose winding number $\nu $
is ``locked'' to the value $w/T$ for all $\Omega $ in a small interval
around $\Omega =w/T$. Wave packets initially trapped in these islands lead to
the QAMs, i.e., a linear growth of $\left\langle \hat{N}\right\rangle _{vT}
\approx avT$, where $a\approx 2\pi (w/T-\Omega)/\epsilon$ \cite{fgr}. For 
$\Omega =w/T$, $a=0$, but in the main-QR limit of $\epsilon \rightarrow 0$ 
the quasiclassical approximation must be replaced by the exact description of 
(\ref{Ubt}) given by the operator (\ref{Ubt1}), with an exponent linear in 
$\hat{N}$. Such an evolution operator corresponds to an integrable classical 
map \cite{mvb}, in contrast with the nonintegrable quasiclassical map for 
$\epsilon \neq 0$, and generally gives a ratchet behavior. Thus, while both a 
QR ratchet (\ref{R}) with $\Omega =w/T$ and a QAM with $\nu =w/T$ exhibit a 
linear growth of $\left\langle \hat{N} \right\rangle _{vT}$, they are basically 
different in nature. However, one may systematically study the quasiclassical 
regime by using the approach introduced in this paper, namely by considering at 
fixed $\eta $ high-order QR ratchets with rational values of $\tau /(2\pi )$ in 
the vicinity of integers.\newline

It interesting to notice that QR ratchets arise, as we have shown, even for {\em 
symmetric} potentials and wave packets, when their symmetry centers do not coincide. 
Using methods similar to those for $\eta =0$ \cite{dd,dd1}, it is easy to 
extend our fixed-$\beta$ results to the general time evolution of the kicked 
particle, involving a superposition of the time evolutions for all $\beta$. 
One then finds that the kicked particle generally exhibits no ratchet 
current for $\tau =2\pi l_{0}$. The QR quadratic behavior for $\eta =0$ is known 
to be robust, under small variations of $\tau $, on some initial time interval 
\cite{kp}. We expect a similar robustness of the $\eta \neq 0$ quantum-resonant
evolution under small variations of $\eta $ and $\tau /(2\pi )$ around their
rational values. Our results should be then realizable in high-precision
experiments such as recent ones \cite{qr0} concerning $\eta =0$ QRs.\newline

This work was partially supported by the Israel Science Foundation (Grant
No. 118/05).\newline


\begin{references}
\bibitem{ao}  M.K. Oberthaler {\it et al.}, Phys. Rev. Lett. {\bf 83}, 4447
(1999).

\bibitem{ao1}  R.M. Godun {\it et al.}, Phys. Rev. A {\bf 62}, 013411
(2000); M.B. d'Arcy {\it et al.}, Phys. Rev. E {\bf 64}, 056233 (2001).

\bibitem{ao2}  S. Schlunk {\it et al.}, Phys. Rev. Lett. {\bf 90}, 054101
(2003); S. Schlunk {\it et al.}, Phys. Rev. Lett. {\bf 90}, 124102 (2003);
Z.-Y. Ma {\it et al.}, Phys. Rev. Lett. {\bf 93}, 164101 (2004); G.
Behin-Aein {\it et al.}, Phys. Rev. Lett. {\bf 97}, 244101 (2006).

\bibitem{fgr}  S. Fishman, I. Guarneri, and L. Rebuzzini, Phys. Rev. Lett. 
{\bf 89}, 084101 (2002); J. Stat. Phys. {\bf 110}, 911 (2003).

\bibitem{qam}  A. Buchleitner {\it et al.}, Phys. Rev. Lett. {\bf 96},
164101 (2006).

\bibitem{qam1}  I. Guarneri, L. Rebuzzini, and S. Fishman, Nonlinearity {\bf
19}, 1141 (2006).

\bibitem{qam2}  R. Hihinashvili {\it et al.}, Physica (Amsterdam) {\bf 226D},
1 (2007).

\bibitem{note}  Since (\ref{Hf}) is translationally invariant in $\hat{x}$,
it can be restricted to the space of Bloch functions $\varphi _{\beta
}(x)=\exp (i\beta x)\psi _{\beta }(x)$ with fixed $\beta $ and arbitrary $
2\pi $-periodic function $\psi _{\beta }(x)$. In fact, $\hat{H}_{\text{f}
}\varphi _{\beta }(x)=\varphi _{\beta }^{\prime }(x)=\exp (i\beta x)\psi
_{\beta }^{\prime }(x)$, where $\psi _{\beta }^{\prime }(x)=\exp (i\beta x)
\hat{H}_{\text{f,}\beta }\psi _{\beta }(x)$ and $\hat{H}_{\text{f,}\beta }$
is given by (\ref{Hf}) with $\hat{p}$ replaced by $\hat{p}+\beta $. One can
interpret $x$ in $2\pi $-periodic functions [such as $V(x),$ $\psi _{\beta
}(x)$, and $\psi _{\beta }^{\prime }(x)$] as an angle $\theta $ and $\hat{p}$
in $\hat{H}_{\text{f,}\beta }$ as an angular-momentum operator $\hat{N}
=-id/d\theta $.

\bibitem{kp}  S. Wimberger, I. Guarneri, and S. Fishman, Nonlinearity 
{\bf 16}, 1381 (2003).

\bibitem{dd}  I. Dana and D.L. Dorofeev, Phys. Rev. E {\bf 73}, 026206
(2006).

\bibitem{dd1}  I. Dana and D.L. Dorofeev, Phys. Rev. E {\bf 74}, 045201(R)
(2006).

\bibitem{qra}  H. Schanz, T. Dittrich, and R. Ketzmerick, Phys. Rev. E {\bf
71}, 026228 (2005); G. Hur, C.E. Creffield, P.H. Jones, and T.S. Monteiro,
Phys. Rev. A {\bf 72}, 013403 (2005); J. Gong and P. Brumer, Phys. Rev.
Lett. {\bf 97}, 240602 (2006).

\bibitem{qrr}  E. Lundh and M. Wallin, Phys. Rev. Lett. {\bf 94}, 110603
(2005); E. Lundh, Phys. Rev. E {\bf 74}, 016212 (2006).

\bibitem{qr}  F.M. Izrailev, Phys. Rep. {\bf 196}, 299 (1990), and
references therein.

\bibitem{cs}  S.-J. Chang and K.-J. Shi, Phys. Rev. A {\bf 34}, 7 (1986).

\bibitem{gs}  B.C. Berndt, R.J. Evans, and K.S. Williams, {\em Gauss and
Jacobi Sums} (John Wiley \& Sons, New York, 1998).

\bibitem{gs1}  See, e.g., Chapter 1 in Ref. \cite{gs}, in particular Sec.
1.5.

\bibitem{js}  The Jacobi symbol $\left( \frac{a}{b}\right) $, where $a$ and 
$b$ are integers and $b$ is positive and odd, is first defined in the case
that $b=l$, a prime number: If $l$ divides $a$ ($l>1$), $\left( \frac{a}{l}
\right) =0$; otherwise, $\left( \frac{a}{l}\right) =1$ if there exists an
integer $d$ such that $l$ divides $a-d^{2}$ and $\left( \frac{a}{l}\right)
=-1$ if such an integer does not exist. Then, if $b$ is a product of $L$
prime numbers not necessarily distinct, $b=l_{1}l_{2}\dots l_{L}$, one
defines $\left( \frac{a}{b}\right) =$ $\left( \frac{a}{l_{1}}\right) \left( 
\frac{a}{l_{2}}\right) \dots \left( \frac{a}{l_{L}}\right) $.

\bibitem{mvb}  M.V. Berry, Physica (Amsterdam) {\bf 10D}, 369 (1984).

\bibitem{qr0}  C. Ryu {\it et al.}, Phys. Rev. Lett. {\bf 96}, 160403 (2006).
\end{references}
\end{document}